\pdfoutput=1

\documentclass[twocolumn,showpacs,preprintnumbers,amsmath,amssymb,superscriptaddress,prl]{revtex4}

\usepackage{graphicx}
\usepackage{dcolumn}
\usepackage{bm}

\begin{document}


\title{ARPES study of the Kitaev Candidate $\alpha$-RuCl$_3$}
\author{Xiaoqing Zhou}
\author{Haoxiang Li}
\author{J.~A.~Waugh}
\author{S.~Parham}
\affiliation{Department of Physics, University of Colorado at Boulder, CO 80309, USA}
\author{Heung-Sik Kim}
\author{J.~A.~Sears}
\author{A.~Gomes}
\affiliation{Department of Physics, University of Toronto, Toronto, ON M5S 1A1, Canada}
\author{ Hae-Young Kee} 
\affiliation{Department of Physics, University of Toronto, Toronto, ON M5S 1A1, Canada}
\affiliation{Canadian Institute for Advanced Research, CIFAR Program in Quantum Materials, Toronto, ON M5G 1Z8, Canada}
\author{ Young-June Kim} 
\affiliation{Department of Physics, University of Toronto, Toronto, ON M5S 1A1, Canada}

\author{D.~S.~Dessau}
\affiliation{Department of Physics, University of Colorado at Boulder, CO 80504, USA}

\date{\today}

\begin{abstract}
$\alpha$-RuCl$_3$ has been hinted as a spin-orbital-assisted Mott insulator in proximity to a Kitaev spin liquid state. Here we present ARPES measurements on single crystal $\alpha$-RuCl$_3$ in both the pristine and electron-doped states, and combine them with LDA+SOC+U calculations performed for the several low-energy competing magnetically ordered states as well as the paramagnetic state. A large Mott gap is found in the measured band structure of the pristine compound that persists to more than 20 times beyond the magnetic ordering temperature, though the paramagnetic calculation shows almost no gap. Upon electron doping, spectral weight is transferred into the gap but the new states still maintain a sizable gap from the Fermi edge. These findings are most consistent with a Mott insulator with a somewhat exotic evolution out of the Mott state with both temperature and doping, likely related to unusually strong spin fluctuations.
\end{abstract}

                                                            
\maketitle

The study of frustrated strongly interacting electronic systems has occupied a forefront seat in condensed matter physics for generations, originating from the separate proposals of the correlation driven insulating state\cite{Mott} and the Resonant-Valence-Band (RVB) model\cite{RVB}. Together, these ideas have spawned many offshoots, for example ideas on quantum criticality\cite{Sachev}, frustrated magnetism\cite{Lee, Balents}, and the spin liquid state\cite{SL1, SL2}, with many of these ideas also connecting to the everlasting puzzle of high temperature superconductivity in cuprates\cite{DopedMott}.  Recently, these ideas have made major inroads into topological quantum computation\cite{QC}, especially when paired with the exactly solvable Kitaev model \cite{Kitaev}. This model is based on spin-1/2 moments arrayed on a honeycomb lattice with a bond-dependent Kitaev coupling term. More generally, considering a system with the Kitaev term $K$ and other spin exchange terms $J$ (such as the usual Heisenberg term) in its Hamiltonian, an increased strength of $K$ relative to $J$ would govern the evolution of magnetically ordered states towards the spin liquid state. It has been proposed that a significant Kitaev term $K$ can be realized in systems with strong spin-orbit coupling effects forming a $J_{eff}=1/2$ state, such as the A$_2$IrO$_3$ iridates\cite{JK1, JK2}. Nevertheless, lattice distortions deviate A$_2$IrO$_3$ iridates from the Kitaev ideal\cite{Ir1, Ir2}, and experimental evidence confirming the ``Kitaev physics" remains scarce and inconclusive.

Recently, $\alpha$-RuCl$_3$\cite{RuCl3a}, a compound with its RuCl$_6$ octahedra sharing edges on a nearly perfect honeycomb lattice, has been highlighted as a perhaps more promising Kitaev candidate, even though its spin-orbital coupling is expected to be weaker. Initially labeled as a semiconductor with a small gap of 0.3 eV\cite{RuCl3b}, it was later discussed as a Mott-insulator\cite{RuCl3c} in which spin-orbit coupling plays a critical role\cite{RuCl3d}. A number of experimental studies with very different techniques have revealed a rich and intriguing paradigm of magnetism in this system\cite{RuCl3e, RuCl3f, RuCl3g, RuCl3h, RuCl3i, neutron, Raman}. In earlier works, an AFM ground state with a Neel temperature $T_N$ of 14 K was suggested by magnetic susceptibility and Mossbauer spectroscopy measurements\cite{RuCl3e, RuCl3f}. Later studies of specific heat, magnetic susceptibility and NMR further identified two distinctive magnetic phases with a pair of magnetic transitions at 8 K and 14 K\cite{RuCl3g, RuCl3h, RuCl3i}. A recent structural study\cite{RuCl3_HBCao} clarify the 7-8 K transition as being associated with a default 3-layer stacking of the C3/2m unit cells in as-grown samples, and the 14 K transition being associated with a 2-layer stacking sequence induced by structural deformation. Neutron scattering studies\cite{RuCl3g, neutron} identified both stacking sequence phase as a ``zig-zag" magnetically ordered state near the spin liquid phase, and have found a collective magnetic excitation mode qualitatively unchanged across $T_N$. The presence of this ordered state raises a question on whether the system is a genuine Mott insulator or a Slater insulator. In parallel, theoretical calculations of RuCl3 band structures have emphasized the importance of the Kitaev exchange interaction to the Mott physics on this system\cite{Kee1, Kee2}, while a direct comparison between theory and experiment has not been established.

In this work, we studied the experimental band structure and density of states of $\alpha$-RuCl$_3$, potassium-intercalated $\alpha$-RuCl$_3$ and \textit{in-situ} Rubidium-doped $\alpha$-RuCl$_3$ with Angle-Resolved Photoemission Spectroscopy (ARPES). At a temperature 30-40 times larger than $T_N$ where long-range magnetic order is absent, we found a large charge gap of 1.2 eV, establishing $\alpha$-RuCl$_3$ as a genuine Mott insulator. Not surprisingly, its band structure cannot be captured by DFT calculations in the paramagnetic state; interestingly, the ARPES spectra resemble a substantially broadened mixture of DFT calculations in several magnetically ordered states in the proximity of a Kitaev spin liquid, suggesting that strong spin fluctuations survive to very high temperatures. Upon electron doping it undergoes a novel Mott transition, in which the Mott gap is abruptly reduced but not closed. These observations possibly highlight the unusual magnetism in the Mott state, which appears to intimately connect to the Kitaev spin liquid phase.
\begin{figure}[t]
\includegraphics[width=85mm]{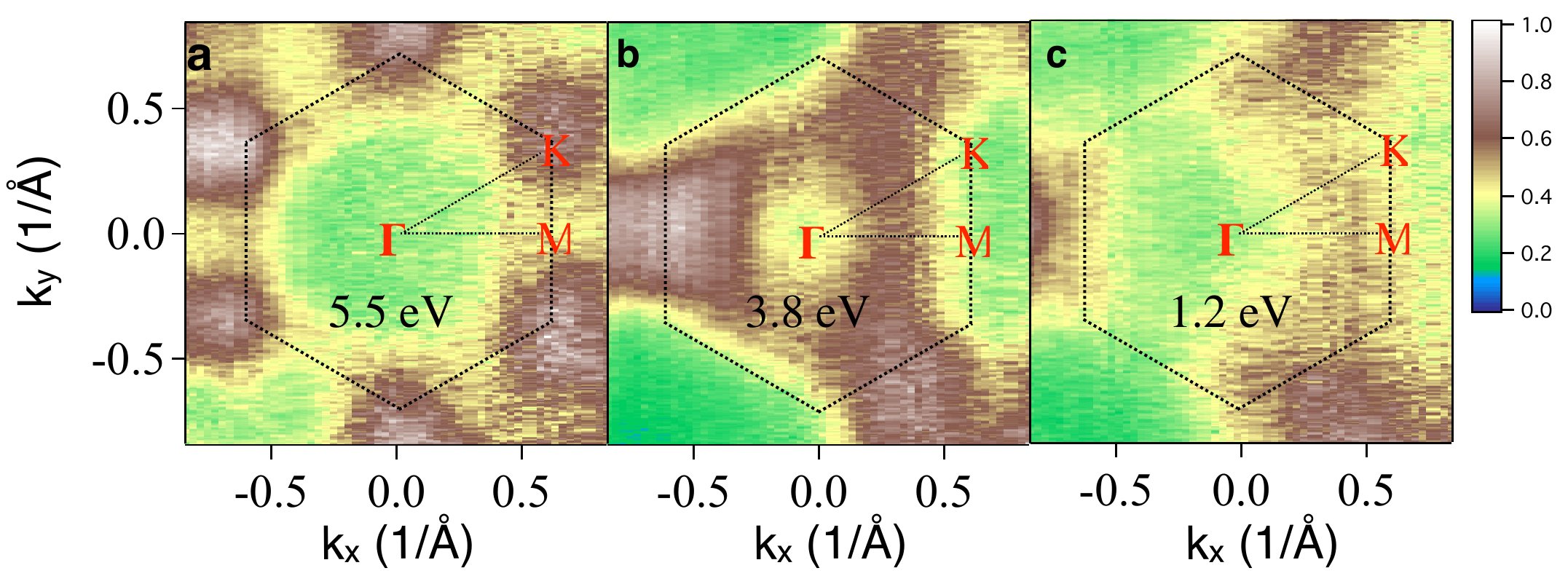}
\caption{\label{fig1}Iso-energy plot for RuCl$_3$ showing a 6-fold symmetry at a) 5.5 eV; b) 3.8 eV and c) 1.2 eV below $E_F$ near the first Brillouin zone. Data were taken in ALS beam line 4.0.3 with $h\nu$ = 75 eV and linear horizontally polarized light.}
\end{figure}
\\

Fig.~1 shows iso-energy plots in momentum space at a variety of binding energies (i.e. energy below $E_F$). A six-fold symmetry is apparent as shown in Figure 1a at binding energy 5.5 eV. Due to the matrix element effect, at different binding energies such as 3.8 eV and 1.2 eV the spectra show six-fold or three-fold symmetry. Such clear symmetries confirm the quality of our sample and spectra.

Figures 2a, 2b and 2c show the rich band structure measured along the high symmetry cut $\Gamma$-M-K-$\Gamma$ under different experimental geometries and incident light polarizations at a temperature of 300 K (see supplementary information). It has been demonstrated\cite{CY} that ARPES data taken with s-polarized light is more sensitive to the in-plane orbitals while the p-polarized light is more sensitive to out-of-plane orbitals. As shown in Figure 2d, this suggests that the bands between -8 to -3 eV (with 0 defined at $E_F$) correspond mostly to in-plane orbitals, in contrast to the bands between -3 to -1 eV which appear to be a mixture of in-plane and out-of-plane orbitals. DFT calculations\cite{Kee1} label the former as Cl p-bands, and the latter Ru t$_{2g}$ bands. Different from the Cl p-bands, the Ru t$_{2g}$ bands can be characterized with very flat dispersion (see supplementary information), anomalously large energy broadening (see below), and most importantly a large charge gap. While ARPES lacks the ability to probe the density of states above $E_F$, with the ``leading edge" method (i.e. measuring the gap from the mid-point of the spectral weight edge) we estimate that the gap is at a minimum 1.2 eV at 300 K.

The existence of such a large gap as well as the giant energy broadening at such a high temperature  has strong implications. First of all, the gap most likely arises from strong spin or charge correlations (i.e. Mott type), since its magnitude cannot be reproduced with LDA-only calculations\cite{Kee1}. More importantly, the gap persists to a temperature 40 times the Neel temperature $T_N$ $\approx$ 7 K, ruling out the possibility of a Slater-type gap which relies on long-range magnetic order. The persistence of the gap at high temperature is reminiscent of the Mott phenomenology in iridates in which spin-orbital coupling plays a vital role\cite{Ir3}, even though the spin orbital coupling effect in RuCl$_3$ is relatively smaller. We note, as show in Figure 2e, our nonmagnetic DFT+SOC+U calculation with a large $U_{eff}$  = 5 eV, yields only a vanishingly small gap of 0.02 eV. To investigate the role of spin correlations played in the Mott physics here, we performed DFT calculations incorporating several magnetically ordered states, as shown by examples of Figure 2f for ``zig-zag" order, Figure 2g for ferromagnetic order, and Figure 2h for antiferromagnetic order, respectively. The qualitative features of the ARPES spectra such as the very flat Ru t$_{2g}$ bands and the magnitude of the gap are well reproduced by the DFT projections, whereas a Gaussian energy broadening of about 0.4 eV has to be introduced in the DFT projections to match the experimental observation (see supplementary information). The broadening shifts spectral weight inside the gap, resulted in an onset of the electron continuum at about 1 eV which is consistent with the optical measurements\cite{Kee2}. In general, we found the ARPES spectra resemble mixtures of these DFT projections, and seems to resemble the one with the zig-zag order most. Please note that while the zig-zag order is the predicted ground state, the ferromagnetic and antiferromagnetic phases are so close in energy ($<$ 4meV) in the extended Kitaev spin model at zero temperature\cite{Kee1} that they would likely become relevant with increasing temperatures. Thus within this Kitaev-type framework, it is natural to find a thermal average of these phase above the Neel temperature, with no single stable, dominant long range magnetic order but instead with fluctuating short-range magnetic correlations, which assist the Mott physics at high temperature. The anomalous energy broadening on the order of 0.4 eV can also naturally arise from this scenario, as the Kitaev spin liquid is notable for hosting very many massless spin excitations. This is quite consistent with the neutron scattering observation of a strong, flat magnetic excitation band above 14 K\cite{neutron}, and the Raman scattering results of a scattering continuum around 100 K\cite{Raman}, even though these temperatures are much lower. In contrast, alternatives interpretations can be mostly ruled out: the instrument resolution is about 20 meV, the sample is rather clean as evidenced by the easy cleavage and clear Cl p-bands, charge fluctuations are forbidden by the insulating gap, and the phonon-related effect should be at least one order of magnitude smaller (about 0.03 eV at 300 K).

\onecolumngrid

\begin{figure}[hbt!]
\includegraphics[width=180mm]{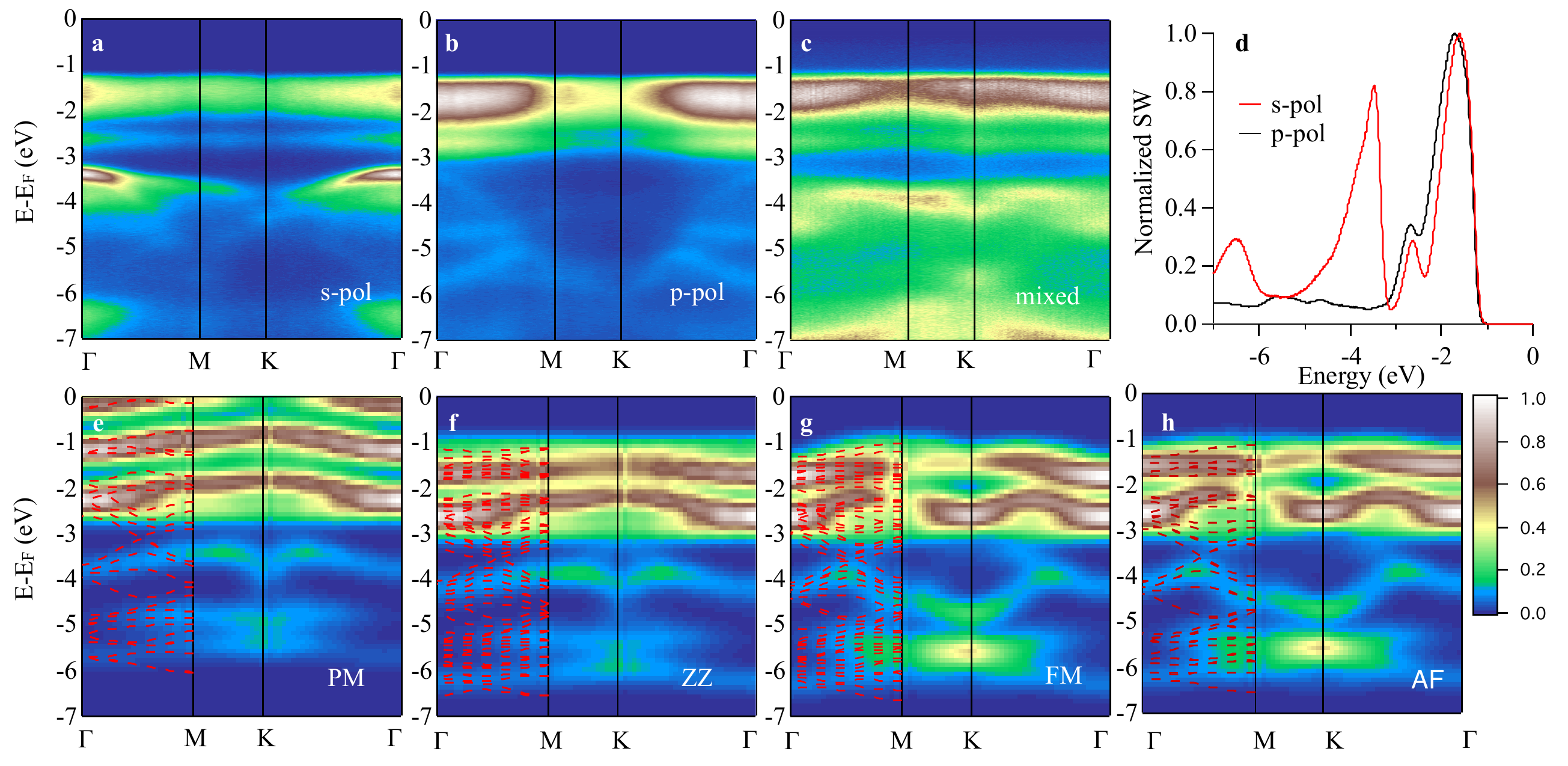}
\caption{\label{fig2}ARPES spectra showing band structures for RuCl$_3$ at a) ALS beam line 10.0.1 with $h\nu$  = 55 eV and s-polarized light; b) ALS beam line 10.0.1 with $h\nu$=55 eV and p-polarized light and c) ALS beam line 4.0.3 with $h\nu$ = 75 eV and linear horizontally polarized light (mixed s and p polarization). d) Comparison of the normalized spectral weight (integrated in k-space) for measurements with a) s-polarized light and b) p-polarized light respectively. e) Broadened DFT calculation with spin-orbital coupling and in the paramagnetic phase; f) with a zig-zag magnetic order; g) with FM magnetic order and h) with AF magnetic order. Band dispersions along $\Gamma$-M direction are represented  by dashed lines.}
\end{figure}

\twocolumngrid

\onecolumngrid

\begin{figure}[hbt!]
\includegraphics[width=170mm]{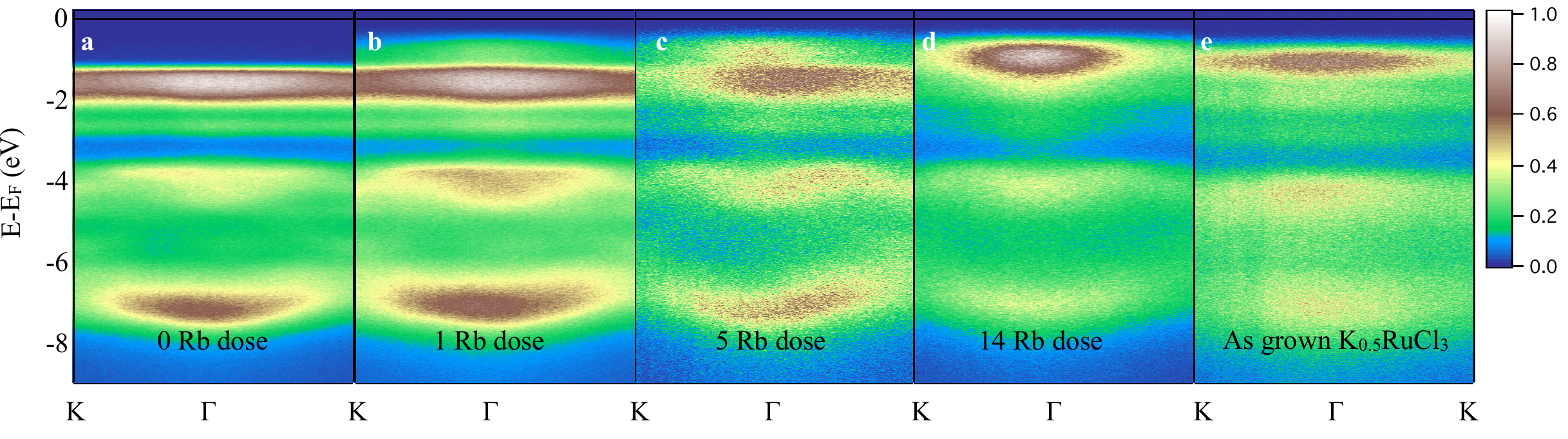}
\caption{\label{fig3}ARPES spectra along $\Gamma$-K direction for RuCl$_3$ with a) no Rb dosing; b) 1 Rb dosing; c) 5 Rb dosing and d) 14 Rb dosing (saturation) at 300 K. Significant weights are shifted into the gap from the Ru valence bands while the Cl bands remain largely unaffected. e) ARPES spectra for as-grown K$_{0.5}$RuCl$_3$ at 300 K. Data were taken in ALS beam line 4.0.3 with $h\nu$ = 75 eV and linear horizontally polarized light. }
\end{figure}
\twocolumngrid

To investigate how the Mott physics and the unusual magnetism evolve with doping, we studied the ARPES spectra of $\alpha$-RuCl$_3$ with gradual \textit{in-situ} rubidium doping (see supplementary information) and as-grown potassium intercalation in ALS beam line 4.0.3 with 75 eV and linear horizontally polarized light. Figure 3a shows the un-doped pristine ARPES spectra along the K-$\Gamma$-K direction. With a single dose of  \textit{in-situ} rubidium doping, it was observed that some spectral weight developed at an energy closer to $E_F$, as shown in Figure 3b. With more dosing, such weight grows significantly, while the Cl valence bands remain qualitatively unchanged. As shown in Figure 3d, with 14 dosings at which the impact of dosing was saturated, the original Ru bands are greatly weakened, indicating that spectral weight has been transferred from the original Mott-like bands to the ``new" bands closer to $E_F$. As indicated by Figure 3e, spectra on as grown potassium-intercalated $\alpha$-RuCl$_3$ show qualitatively the same behavior. Even though there is a lack of clear dispersion possibly due to disorder, the locations of all the spectral weight maxima can be approximately tracked to their counterparts in Figure 3d. This indicates that the evolution we observed with in-situ Rubidium-dosing is very similar to the intrinsic and bulk physics in the intercalated system.  

\begin{figure}[t]
\includegraphics[width=85mm]{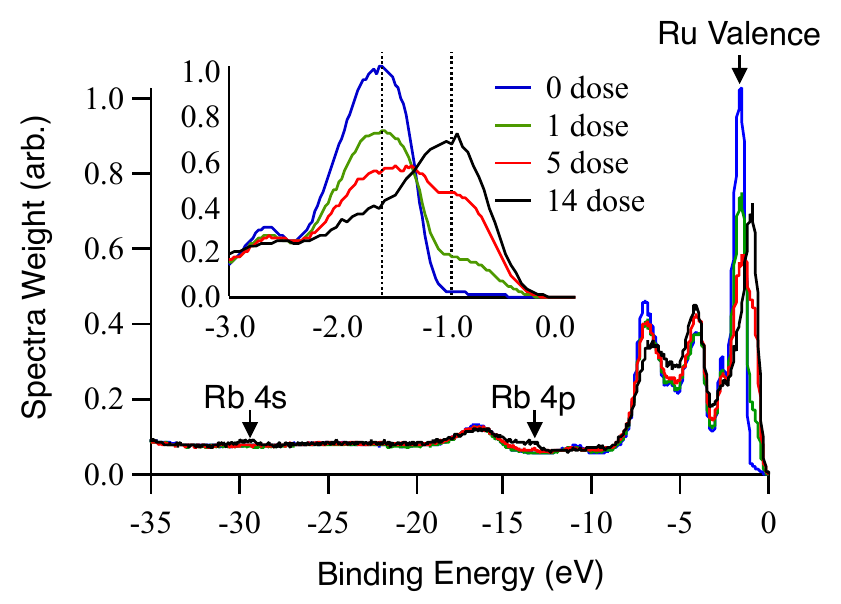}
\caption{\label{fig4} a) Core level scan of RuCl3 with in-situ Rb dosing. Inset: spectral weight transfer into the gap (zoomed in near EF).}
\end{figure}

To further illustrate the evolution with doping, we compared the integrated spectral weights in Figure 4 at different dosing levels. In principle, the integrated weight is proportional to the density of states that is associated with occupied bands. We found that the small amount of Rubidium dosed on the sample surface had very little impact on the spectral weight at deep binding energies ($>$10 eV). The chemical potential, as referenced by the many peak positions, shows minimal change even with 14 doses. Such a phenomenology is in sharp contrast to a similar study on another Kitaev candidate Na$_2$IrO$_3$, in which the chemical potential is gradually shifted forward with K-dosing \cite{Ir3}. There is a small suppression of the spectral weight of the Cl bands from 3 to 8 eV binding energy, which can be readily understood as the attached Rubidium atoms scattering the photoelectrons. The most drastic change happens near $E_F$, where the spectral weight of the original Ru bands is substantially suppressed, along with a large increase in spectral weights peaking at 0.9 eV. To first order, there seems to be weight transfer between the original bands and the ``new" bands. In addition, with substantial Rb-dosings we managed to lessen the impact of sample charging, and were able to approach ARPES spectra at lower temperatures. Asides from the persisting charging effect, we found there was practically no change in the electronic structure from 300 K to 100 K (see supplementary information). Such a continuity, along with the scattering continuum from 100 K to 5 K as observed in the Raman scattering study\cite{Raman}, favor the scenario that spin fluctuations persist to much higher than the Neel temperature.

\begin{figure}[t]
\includegraphics[width=85mm]{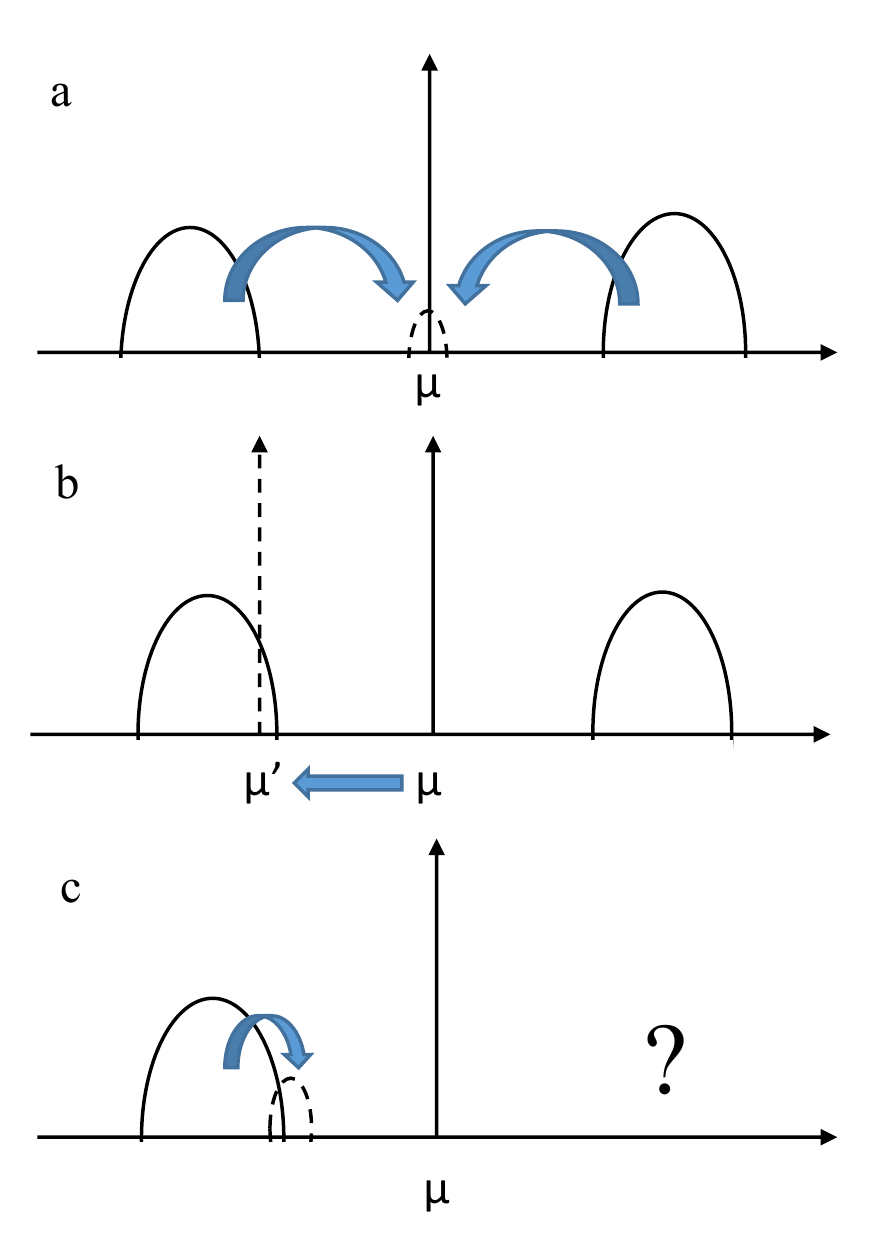}
\caption{\label{Mott} a) Sketch of an ordinary Mott transition, in which the spectral weights (solid line) is transferred to the new states (dashed line) at middle of the Mott gap at the chemical potential $\mu$. b) Sketch of another ordinary Mott transition, in which the chemical potential is lowered towards the lower Hubbard band. c) Sketch of the Mott transition observed in this study, in which the spectral weights (solid line) is transferred to new states (dashed line) that remains gapped from the chemical potential.}
\end{figure}

These observations suggest a novel type of Mott transition. Conventionally, the Mott transition is often thought to happen in two ways, either that spectral weight is abruptly filled in the middle of the Mott gap\cite{MottTransition}, or that the chemical potential abruptly lowers towards the lower Hubbard bands, as indicated by the sketches in Fig.~\ref{Mott}a and \ref{Mott}b. With the Mott transition, the system generally evolves from an insulator to a metal. In this study, we did not observe any notable chemical potential shift due to the electron doping, as evidenced by the stable location of the Cl bands as a function of dosing level. While we did observe an abrupt spectral weight transfer from the 1.6 eV peak to the 0.9 eV peak as soon as we doped the system, the ``new" state is clearly not a metal even with substantial doping, as none of these new states reach $E_F$, as shown in the sketch of Fig.\ref{Mott}c. Considering that in this system  the persistence of the gap with electron doping suggests a persistence (but weakening) of the spin correlations with electron doping, it is likely that the spin fluctuations play an important role in the Mott transition.  

\begin{figure}[t]
\includegraphics[width=60mm]{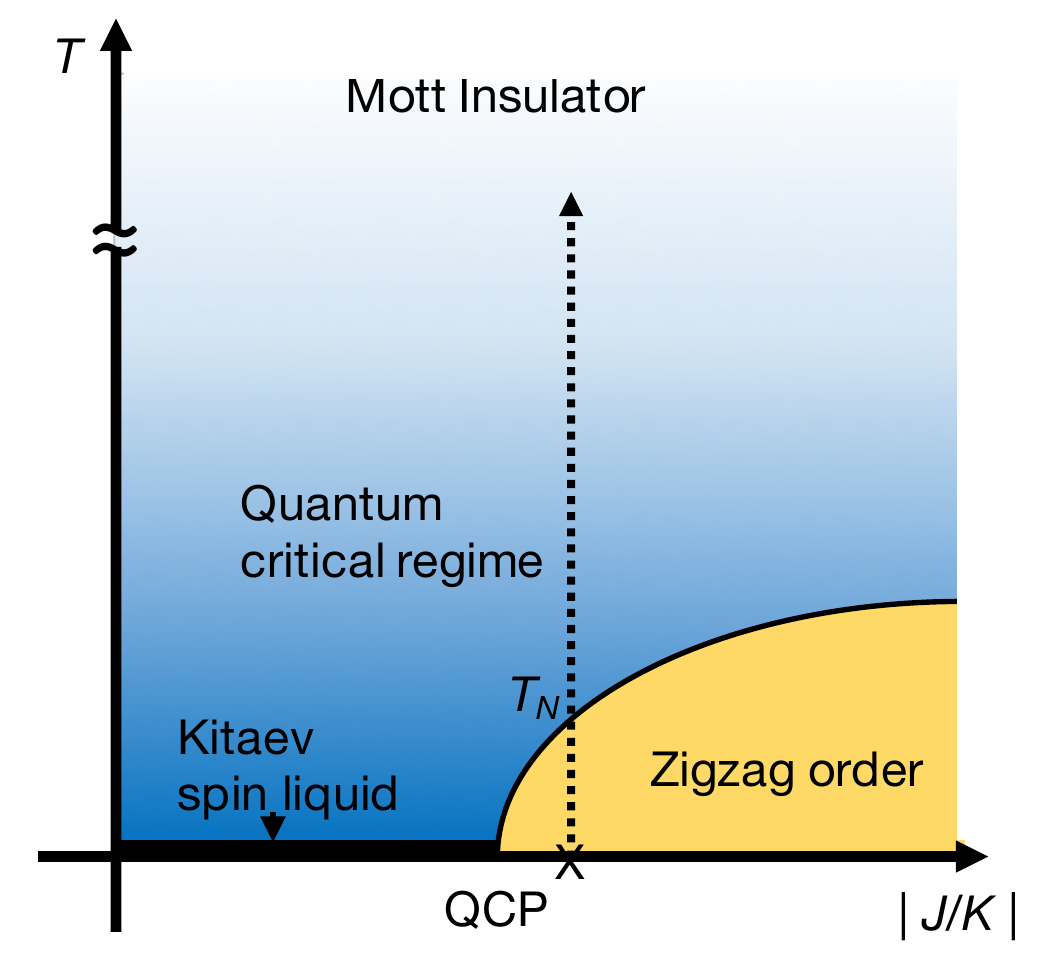}
\caption{\label{fig5} RuCl$_3$ (cross) in the potential phase diagram as described by the $J$-$K$-$\Gamma$ model[26]. The competition between the Kitaev term $K$ and other spin exchange terms $J$ results in a continuous quantum critical area above Neel temperature in which our ARPES observations were made.}
\end{figure}

Overall, the persistence of the gapped band structure across a large temperature range and upon surface-dosing, support $\alpha$-RuCl$_3$ as a Mott insulator with strong spin fluctuations far above $T_N$. Such a phenomenology can be consistently understood within the extended Kitaev spin liquid framework\cite{Kee1}, as shown in Figure 5. The general agreement between the ARPES spectra and the DFT calculations suggests the presence of a substantial Kitaev term $K$ in $\alpha$-RuCl$_3$, which locates it in the vicinity of the Kitaev spin liquid phase. The Kitaev spin liquid, due to its unique property of supporting continuous gapless spin excitations upon quantum and thermal fluctuations, might give rise to a quantum critical regime in which the intriguing Mott physics far above the Neel temperature manifests. 

As a summary, in this work we present a thorough ARPES study of $\alpha$-RuCl$_3$, potassium-intercalated $\alpha$-RuCl$_3$ and \textit{in-situ} Rubidium-doped $\alpha$-RuCl$_3$. At a temperature much higher than the Neel temperature, the band structure shows a large Mott gap and a giant energy broadening that appears only possible with the assistance of surviving short-range magnetic correlations. With electron doping this gap is abruptly reduced but not destroyed, which we interpret as a novel Mott transition driven by the evolution of magnetic correlations. These results are consistent with $\alpha$-RuCl$_3$ being in the proximity of a Kitaev test-ground, with rich evolutions of magnetic correlations and fluctuations towards the spin liquid.
\\

\noindent\textbf{Acknowledgement}. We acknowledge Drs. Y. D. Chuang, J. D. Denlinger, and S. K. Mo for technical assistance. X.Q. Zhou thanks Dr. B. Normand for valuable discussions. Financial assistance is from the US Department of Energy under grant DE-FG0203ER46066. Research at the University of Toronto was supported by the Natural Science and Engineering Research Council of Canada, Canada Foundation for Innovation, Ontario Ministry of Research and Innovation, and Canada Research Chair program. The experiments were performed at beamlines 4.0.3 and 10.0.1 of the Advanced Light Source, Berkeley, which is supported by the US Department of Energy. Correspondence and requests for materials should be addressed to D.S.D. (Dessau@Colorado.edu)

\end{document}


\title{Supplementary information: ARPES study of the Kitaev Candidate $\alpha$-RuCl$_3$}



%

\maketitle

\onecolumngrid

\title{Supplementary information: ARPES study of the Kitaev Candidate $\alpha$-RuCl$_3$}

\maketitle

\section{Sample Preparation and Experimental Setup}

Single crystal samples of $\alpha$-RuCl$_3$ were prepared by vacuum sublimation from commercial RuCl$_3$ powder in sealed quartz tubes. The magnetic susceptibility and specific heat data of such samples can be found in reference [1]. In particular, neutron diffraction 
measurements showed the crystal structure at low temperature to have a predominantly three-layer stacking periodicity, with stacking faults resulting in a minor phase with two-layer periodicity.

The electrochemical doping was accomplished by mounting the crystal in a 
platinum sample holder and doping galvanostatically in a solution of 
KNO3. Preliminary structural measurements show that the chemical 
intercalation of water and K+ ions increases the c lattice parameter by 
50\%, a result in general agreement with the findings in [2].

Experimental ARPES measurements were carried out at the Advanced Light Source (ALS) beamlines 10.0.1 and 4.0.3. Data from beamline 10.0.1 were taken with a glancing incident light and with accurate control of light polarization relative to the sample plane. Such a setup allows for a deduction of the orbital symmetry from the ARPES matrix elements. Beamline 4.0.3 allows an \textit{in-situ} Rubidium doping of the Mott insulator. Unless stated otherwise, all data were taken at a temperature of 300 K without the complication of the sample charging problem, which becomes more and more problematic with decreasing temperature.  Specifically, Figure 2a corresponds to data taken at beam line 10.0.1 with 55 eV photon energy and with 100 percent s-polarized light, while Figure 2b with mostly (97 percent) p-polarized light. The beam line 4.0.3 data, as shown in Figure 2c with photon energy of 75 eV, shall be considered as a supposition of these two cases.

\section{DFT calculations}

DFT calculations in this study are obtained by using the Vienna ab-initio Simulation Package (VASP) [3, 4]. For the calculation we used a unit cell with an isolated RuCl$_3$ honeycomb layer with 15$\textrm{\AA}$ thick vacuum between the adjacent layers. The single-layer structure, with the in-plane lattice constant a = 5.98$\textrm{\AA}$ [5], was obtained from the structure optimization with a force criterion of 1 meV/$\textrm{\AA}$. A revised Perdew-Burke-Ernzerhof generalized gradient approximation (PBEsol)[6] was used. 15×15 Monkhorst-Pack grid and 370 eV of plane-wave energy cutoff were used for the momentum space sampling and the basis set, respectively. For the DFT+U +SOC calculations we employed Dudarev's rotationally invariant formalism [7] with effective $U_{eff} = U - J $= 5 eV. When simulating ARPES spectra, we assumed that the Fermi level is at the center of the band gap, and the electric field is perpendicular to the Ru honeycomb layer.

\section{Band dispersions in ARPES spectra}

A notable feature in our ARPES data in figures 2a, 2b and 2c is that the Ru t$_{2g}$ bands from -3 eV to -1 eV appear to be quite broad in energy. For a better comparison with the DFT calculations of dispersions, we use the ``curvature method" [8] to identify the main band dispersions from the ARPES spectra. The curvature method narrows down peak widths from the local maxima in the EDC (energy-dependent curves) of the original ARPES spectra (shown in Figure S1a, S1b and S1c). With the curvature method we can identify very flat Ru bands reminiscent of those in the DFT calculations, as shown in Figure S1d, S1e and S1f. These can be compared with the DFT calculations without the Gaussian energy broadening, as shown in Figure S1g, S1h and S1i for DFT calculations with the ZZ (zig-zag) order, the ferromagnetic order and the antiferromagnetic order respectively. On the other hand, one can impose a Gaussian convolution in energy on the curvature method plots to qualitatively reproduce the raw ARPES spectra, from which an energy broaden on the order of 0.4 eV or more can be estimated.

\begin{figure}[hbt!]
\includegraphics[width=170mm]{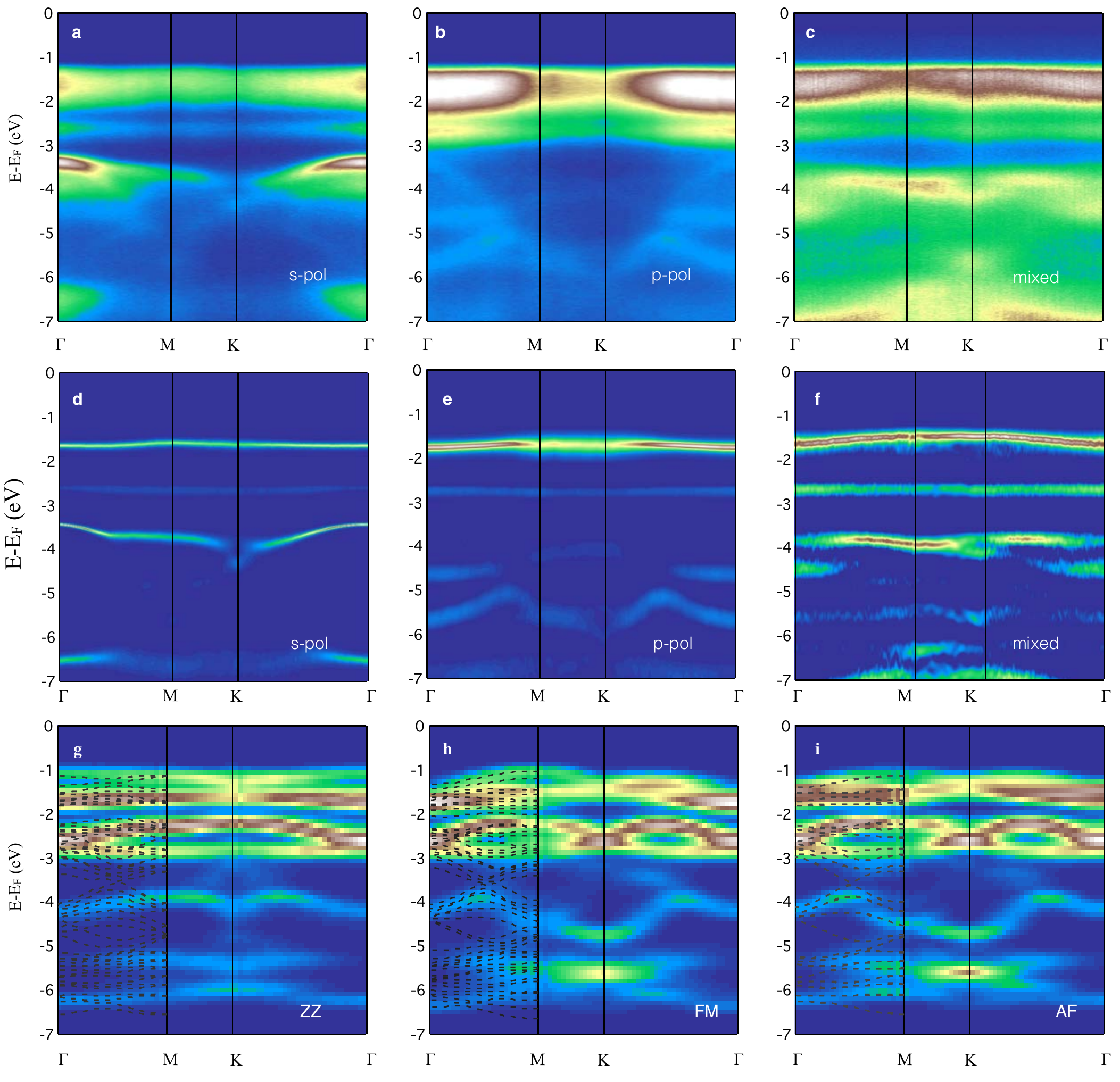}
\caption{\label{figS1}ARPES spectra showing band structures for RuCl$_3$ at a) ALS beam line 10.0.1 with $h\nu$  = 55 eV and s-polarized light; b) ALS beam line 10.0.1 with $h\nu$=55 eV and p-polarized light and c) ALS beam line 4.0.3 with $h\nu$ = 75 eV and linear horizontally polarized light (mixed s and p polarization). d) e) f) Band dispersions obtained with ``curvature method"[8] for panel a) b) and c) respectively. These dispersions can be compared with DFT calculations in g) the zig-zag magnetically ordered phase; h) the ferromagnetic phase and i) antiferromagnetic phase.}
\end{figure}

\section{$\textbf{Rb}$-dosing towards low temperature}

Rubidium doping was achieved by evaporating a Rubidium filament towards to the sample. It took the form of consecutive doses, with one dose corresponding to a filament heating  current of about 5A that lasts 60 seconds. As mentioned in the manuscript, with substantial Rb-dosing we were able to measure ARPES spectra at temperature lower than 300 K. Figure S2 show the examples of ARPES spectra at 300 K, 180 K, 130 K and 100 K respectively. At 130 K the impact of Rb-dosing had already been saturated, and there is a charging effect that shifts the chemical potential by 0.3 eV. At 100 K the charging effect became stronger, and a 3.7 eV shift in chemical potential was estimated.

Aside from the chemical potential shift, the spectra at these temperatures show little differences except for the weight transfer near $E_F$ as discussed in the manuscript. The physics that governs the band structure is thus approximately temperature insensitive, even though certain anomalies have been identified in the bulk susceptibility within this temperature range. On the other hand, such an observation is consistent with the spin liquid scenario that has no true phase transition within the quantum critical regime as shown in Figure 5.

\begin{figure}[hbt!]
\includegraphics[width=170mm]{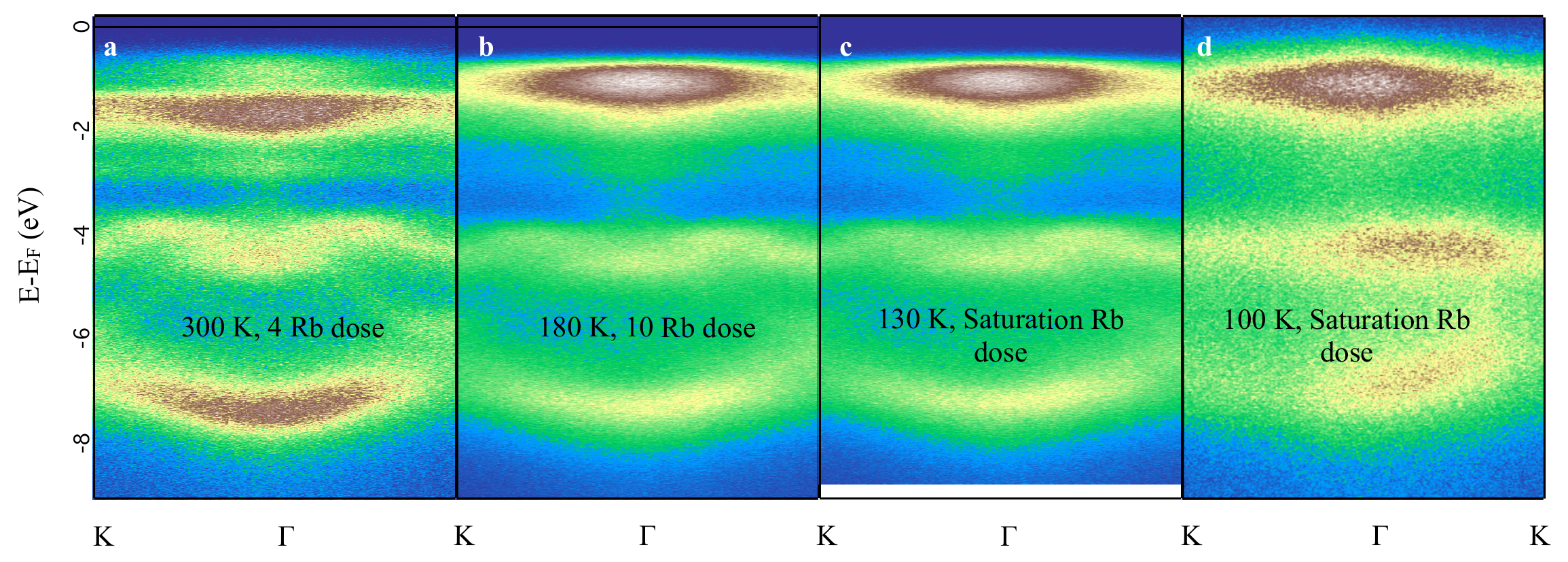}
\caption{\label{figS2}ARPES spectra along $\Gamma$-K direction for a RuCl$_3$ sample with a) 300 K and 4 Rb dosing; b) 180 K and 10 Rb dosing; c) 130 K and 14 Rb dosing (saturation) and d) 100 K and 14 Rb dosing. The data has been corrected for  a 0.3 eV and 3.7 eV shift in work function for c) and d) due to charging of the sample.}
\end{figure}